\documentclass{article}
\usepackage{amsmath}
\usepackage{amssymb}
\usepackage{graphicx}
\usepackage{hyperref}
\usepackage{geometry}
\usepackage{color}
\usepackage{tabularray}
\usepackage{array}
\usepackage{bigstrut}
\usepackage{multirow}
\usepackage{bm}
\usepackage{booktabs}
\usepackage[numbers]{natbib}

\begin{document}

\title{PCAC-GAN: A Sparse-Tensor-Based Generative Adversarial Network for 3D Point Cloud Attribute Compression}
\author{Xiaolong Mao, Hui Yuan, Xin Lu, Raouf Hamzaoui and Wei Gao}
\date{}
\maketitle

\begin{abstract}
 Learning-based methods have proven successful in compressing  geometric information for point clouds. For attribute compression, however, they still lag behind non-learning-based methods such as the MPEG G-PCC standard. To bridge this gap, we propose a novel deep learning-based point cloud attribute compression method that uses a generative adversarial network (GAN) with sparse convolution layers. Our method also includes a module that adaptively selects the resolution of the voxels used to voxelize the input point cloud. Sparse vectors are used to represent the voxelized point cloud, and sparse convolutions process the sparse tensors, ensuring computational efficiency. To the best of our knowledge, this is the first application of GANs to compress point cloud attributes. Our experimental results show that our method outperforms existing learning-based techniques and rivals the latest G-PCC test model (TMC13v23) in terms of visual quality.
\end{abstract}

\section{Introduction}
Point cloud data, which consists of a collection of points in three-dimensional (3D) space, has become increasingly popular for representing scenes and objects. Each point in a point cloud is defined not only by its spatial coordinates but may also include additional attributes such as color and reflectance. These characteristics give point clouds extensive applicability across multiple domains. For instance, in virtual reality (VR) and augmented reality (AR), point clouds are used to create highly realistic 3D models, offering users an immersive experience. In the field of autonomous driving, LiDAR-generated point clouds are crucial for the safe navigation of vehicles. In urban planning and architectural design, point clouds facilitate efficient 3D modeling of urban environments. Additionally, in the conservation of cultural heritage, point cloud technology plays a pivotal role in the digital documentation and preservation of historical sites and artifacts. Point cloud compression is crucial for efficient  point cloud data usage. Point cloud compression aims to reduce storage space requirements and to accelerate processing and transmission speeds, while maintaining the richness of information and detail in  point cloud data, thereby supporting applications across various fields.

There are two primary types of point cloud compression: geometry compression and attribute compression. Geometry compression concerns compressing the 3D coordinates of the points \cite{1}, while attribute compression aims to minimize the redundancy among point attributes \cite{4,5,6}. This paper specifically addresses attribute compression, assuming that the geometry has already been encoded losslessly. It also assumes that the point attributes are given by  color information in  YUV color space. 

\begin{table*}
\centering
\caption{Classification of point cloud attribute compression methods}
\begin{tabular}{|l|l|} 
\hline
Category                               & Method                                                             \\ 
\hline
\multirow{12}{*}{Transform-based PCAC} & Haar Wavelet transform-based algorithm \cite{30}                         \\
                                       & Combination of predictive coding and Haar Wavelet transform \cite{31}    \\
                                       & Graph Fourier transform \cite{9}                                        \\
                                       & Combination of graph transform and discrete cosine transform \cite{32}   \\
                                       & Combination of block-based prediction and graph transform \cite{33}      \\
                                       & Three fine-grained correlation representations \cite{34}                 \\
                                       & Combination of 3D-block-based prediction and transform coding \cite{35}  \\
                                       & Graph Fourier transform based on normalized graph Laplacian \cite{36}    \\
                                       & Combination of Hierarchical transform and arithmetic coding \cite{7}     \\
                                       & Graph transform with optimized Laplacian sparsity \cite{37}              \\
                                       & Stationary Gaussian process\cite{38}                                    \\
                                       & Joint optimized graph transform and entropy coding \cite{48}             \\ 
\hline
\multirow{6}{*}{Learning-based PCAC}   & Sparse-PCAC \cite{26}                                                    \\
                                       & Deep-PCAC \cite{28}                                                      \\
                                       & Folding-based compression~of point cloud attribute \cite{29}             \\
                                       & Lvac \cite{39}                                                           \\
                                       & Tree structure initial encoding \cite{40}                                \\
                                       & CARNet \cite{41}                                                         \\
\hline
\end{tabular}
\label{related work}
\end{table*}

Conventional point cloud attribute compression methods can be broadly categorized into three types: transform-based approaches~\cite{7,8,9}, distance-based methods \cite{10,11,12}, and projection-based methods~\cite{13,14}. Transform-based methods apply geometry dependent transforms to compress attributes, while distance-based methods compress attributes from a coarse to a fine level of detail. Projection-based methods convert point clouds into images or videos and apply existing image/video codecs (e.g., JPEG \cite{15} and H.265 \cite{16}) to achieve compression.

Deep neural networks (DNNs) have gained increasing popularity in academic and industrial contexts for learning-based point cloud compression, largely because of their success in image and video compression \cite{17,18,19}. Most learning-based point cloud compression methods focus on compressing point cloud geometry occupancy using 3D representation models, such as 3D voxel grids \cite{20,21,22}, octrees \cite{23,24}, and sparse tensors \cite{25,26}. These methods outperform traditional rule-based methods like G-PCC \cite{27} in terms of coding performance, due to DNNs’ powerful representation capacity.

Some powerful models, such as PointNet-style architectures \cite{28} and neural network-based 3D to 2D projection \cite{29}, have been successfully used to compress attributes. However,  existing learning-based methods are still not as effective as G-PCC, the current state-of-the-art method.

To address this challenge, we propose a novel learning-based approach called PCAC-GAN. Our method uses a GAN that consists of sparse convolution layers to compress 3D point cloud attributes. We first use a pre-trained neural network module to adaptively voxelize the input raw point cloud at two voxel resolutions based on the density of the point cloud data. Then, we represent the color attributes of the point cloud using sparse tensors and construct a neural network using sparse convolutions. The encoder stacks sparse convolution layers to encode the attributes, while the decoder uses transposed sparse layers to decode them. This method effectively exploits the sparsity of voxelized point clouds and addresses the issue of high complexity and low efficiency that arises from GANs.

One critical rationale for incorporating GANs into the compression of point cloud attributes lies in the distinctive nature of GANs as generative models, which distinguishes them from the reconstruction models commonly used in traditional compression techniques. Unlike conventional approaches that primarily focus on reconstructing compressed representations to recover the original data, GANs adopt a unique strategy by generating novel data that closely resembles the original content. This unique characteristic makes GANs highly suitable for point cloud attribute compression, enabling the recovery of attribute information that may be lost or distorted during the preprocessing and quantization stages.

This paper presents several contributions to the compression of point cloud attributes:
\begin{itemize}

\item We propose a novel approach for compressing 3D point cloud attributes using a GAN consisting of sparse convolution layers. To the best of our knowledge, this is the first time GANs have been applied to point cloud attribute compression.

\item We propose a novel multi-scale transposed sparse convolutional decoder that contributes to achieving a higher compression quality and ratio in learning-based compression systems at reasonable computational effort.

\item We develop an adaptive voxel resolution partitioning module (AVRPM) to partition the input point cloud into blocks with adaptive voxel resolutions. This feature enables AVRPM to effectively process point cloud data with varying densities while maintaining high accuracy and stability.
\end{itemize}
Overall, the proposed PCAC-GAN model shows considerable potential for improving the efficiency of point cloud attribute compression. Our innovative use of GANs and sparse convolution layers may open up new possibilities for tackling the challenges associated with point cloud attribute compression.

\begin{figure*}
 \centering
 \includegraphics[width=\textwidth]{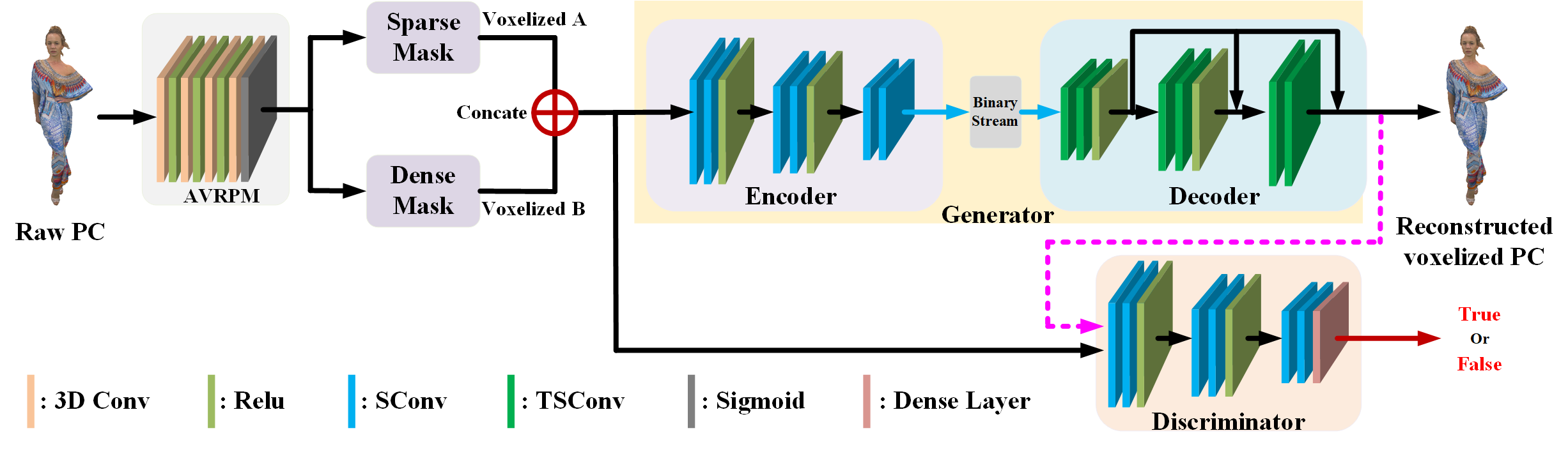}
 \caption{PCAC-GAN architecture. The network consists of AVRPM and a GAN. `Concate' means concatenating two voxelized point clouds with different resolutions. `SConv' and `TSConv' stand for sparse convolution and transposed sparse convolution, respectively.}
 \label{fig1}
\end{figure*}

\section{Related Work}
Our literature review focuses on progress made in two areas: transform-based and learning-based methods for point cloud attribute compression. Transform-based methods rely heavily on complex transform operations, whereas learning-based methods use deep learning models. Table~\ref{related work} presents a classification of  state-of-the-art methods.

\subsection{Transform-based Point Cloud Attribute Compression}
 In the field of point cloud attribute compression (PCAC), various transform-based methods have been proposed. One commonly used approach is the range adaptive Haar transform (RAHT)-based method, which is used in MPEG G-PCC \cite{27} and TMC13 \cite{7}. Zhang et al. \cite{30} proposed a Haar Wavelet transform-based algorithm that takes the surface orientation of the point cloud into account. Chen et al. \cite{31} introduced a combination of predictive coding and the Haar wavelet transform tree specifically for LIDAR point clouds. Another category of methods relies on graph signal processing techniques. Zhang et al. \cite{9} used a graph Fourier transform, while Liu et al. \cite{32} presented a graph transform and discrete cosine transform-based approach. Robert et al. \cite{33} proposed a method that incorporates block-based prediction and graph transforms. Song et al. \cite{34} introduced three fine-grained correlation representations, and Cohen et al. \cite{35} developed a 3D-block-based prediction and transform coding method. Chen et al. \cite{36} designed a weighted graph using a self-loop and defined a graph Fourier transform based on the normalized graph Laplacian. Ricardo et al. \cite{7} used a hierarchical transform and arithmetic coding. Shao et al. \cite{37} introduced a binary tree-based point cloud partitioning technique and explored the graph transform with optimized Laplacian sparsity. Additionally, Ricardo et al. \cite{38} used a stationary Gaussian process to model the statistics of the signal on points in a point cloud. Gao et al. \cite{48} proposed a jointly optimized graph transform and entropy coding scheme that achieves excellent results for compressing point cloud color attributes. However, while transform-based compression methods for point cloud attributes have shown promise, they often come with high encoder complexity due to the intricate transformation operations required for data compression. Moreover, some of these methods rely on prior assumptions about the point cloud data or require specific encoding strategies, which limits their versatility and robustness in practical applications.

\subsection{Learning-based PCAC}
 Learning-based PCAC techniques leverage machine learning approaches to compress point cloud attributes. Wang et al. \cite{26} developed a variational autoencoder (VAE) framework that uses sparse tensors to compress color attributes. Sheng et al. \cite{28} proposed an end-to-end deep lossy compression framework called Deep-PCAC, which directly encodes and decodes attributes without the need for voxelization or point projection. Quach et al. \cite{29} explored folding-based compression techniques that fold a 2D grid onto point clouds and use an optimization mapping method to map point cloud attributes onto the folded 2D grid. Isik et al. \cite{39} focused on compressing the parameters of the volume function and used a coordinate-based neural network to represent the function within each block. Fang et al. \cite{40} introduced an efficient initial encoding method that decomposes point cloud attributes into low-frequency and high-frequency coefficients and models them using a tree structure. Ding et al. \cite{41} proposed a learning-based adaptive loop filter to mitigate compression artifacts in point cloud data, thereby facilitating point cloud attribute compression. Although learning-based PCAC methods have demonstrated impressive results in practical applications by achieving high reconstruction quality and preserving information accuracy while maintaining compression rates, a performance gap still exists when compared to state-of-the-art G-PCC attribute compression techniques. This gap motivates the exploration of novel methodologies and improvements in learning-based PCAC to bridge the performance disparity and further advance the field of point cloud attribute compression.

\section{Proposed Method}
We assume that the point cloud geometry data has already been encoded using a geometry codec. The geometry data is then used as auxiliary information to support attribute compression. The proposed PCAC-GAN framework,  illustrated in Fig.~\ref{fig1}, leverages the adaptability of GANs to learn the input data distribution and to better handle noise and variation in point cloud data. GANs have the ability to capture the complex relationships in point cloud attributes and generate similar point cloud attribute information to the original data, thus minimizing the distortion in the compressed data. The framework consists of two modules: AVRPM and a GAN module for encoding and decoding.

\subsection{AVRPM}
3D point cloud data is  generated by diverse sensors and acquisition methods, leading to variations in density, distribution, and noise characteristics. As point cloud data typically consists of a substantial number of points, it is crucial to consider both efficiency and quality during data processing. However, when dealing with point cloud data that exhibits non-uniform density, using a fixed resolution for voxelization across the entire point cloud may result in the loss of detailed information in specific regions.

To overcome these challenges, the proposed AVRPM intelligently partitions the point cloud into blocks with different densities and chooses appropriate voxel resolutions for each block. In our experiments, we used a voxel resolution of 8×8×8 for blocks with relatively low density and 16×16×16 for blocks with relatively high density. This partitioning strategy allows for parallel processing, optimizing the efficiency of our method.

The primary objective of AVRPM is to enhance the accuracy and precision of point cloud processing while mitigating the imbalanced processing between dense and sparse blocks. By adapting the voxel resolution based on the local density variations, we ensure that each block is processed with optimal efficiency and without compromising the quality of the results. Furthermore, the parallel processing capability of our method guarantees efficient data processing without significantly increasing the overall encoding and decoding pre-processing time.

By effectively addressing the challenges posed by non-uniform density in point cloud data and incorporating parallel processing, AVRPM offers an improved approach for point cloud attribute compression, enabling accurate and efficient processing of diverse point clouds.

The proposed adaptive voxelization module for point clouds applies block segmentation on the raw input point cloud, producing a binary mask that labels points as belonging to either relatively sparse or dense blocks. The module comprises multiple layers of 3D convolutional layers and activation functions, culminating in a sigmoid function that outputs two point cloud masks. Following pre-training, the raw input point cloud is segmented into denser and sparser portions using these two masks.

The loss function of AVRPM consists of two losses: one for supervising the sparse block mask and the other for supervising the dense block mask. The losses are based on the binary cross-entropy (BCE) loss:
\begin{equation}
\text{BCE}(\textbf{y},\boldsymbol{\hat{y}})=-\textbf{y}\log(\boldsymbol{\hat{y}})-(1-\textbf{y})\log(1-\boldsymbol{\hat{y}}),
\end{equation}
where \textbf{y} represents the ground truth label and \bm{$\hat{y}$ } indicates the predicted label generated by the model.

The loss function for the sparse mask is:
\begin{equation}
  \text{loss}_{\text{sparse}}=\text{BCE}(\textbf{M}_\text{sparse},\textbf{G}_{\text{sparse}}),
\end{equation}
where $\textbf{M}_{\text{sparse}}$ and $\textbf{G}_{\text{sparse}}$ represent the ground truth sparse mask and predicted mask, respectively. Similarly, the loss function for the dense mask is 
\begin{equation}
  \text{loss}_{\text{dense}}=\text{BCE}(\textbf{M}_\text{dense},\textbf{G}_{\text{dense}}),
\end{equation}
where $\textbf{M}_{\text{dense}}$ and $\textbf{G}_{\text{dense}}$ represent the ground truth dense mask and predicted mask, respectively. The overall loss function is
\begin{equation}
  \text{loss}_{\text{total}}=\alpha\mathrm{loss}_{\mathrm{sparse}}+(1-\alpha)\mathrm{loss}_{\mathrm{dense}},
\end{equation}
where $\alpha$ is used to balance the two loss functions. Here, we set $\alpha$ to 0.3 based on experience, as this value can emphasize the accuracy of sparse mask encoding.

\subsection{Encoder and Decoder}
The proposed method introduces a novel approach that combines the power of autoencoder and GAN architectures, resulting in the generation of high-quality point clouds while maintaining model stability within reasonable constraints. In addition to these architectural advances, the method capitalizes on sparse tensors and sparse convolution techniques to achieve efficient feature representation and processing. 

A sparse tensor is a specialized data structure designed to store and process high-dimensional data sets that contain a significant number of zero elements. It efficiently represents sparse data by only storing non-zero values along with their corresponding indices, effectively reducing memory consumption and computational requirements. By leveraging sparse tensors for point cloud data representation, several advantages are attained, including high storage efficiency, computational efficiency, and strong compression capabilities. These benefits arise due to the abundance of zero elements in the sparse tensor representation.

In the context of point clouds, the input data typically consists of both geometry coordinates (\textbf{$\Vec{C}$}) and attribute information (\textbf{$\Vec{F}$}). However, the focus of our paper is on compressing the attribute information (denoted by \textbf{$\Vec{F}$}) and reconstructing it (denoted by $\hat{\textbf{F}}$) from its compressed form. By leveraging sparse tensor techniques, we can efficiently compress the attribute information, reducing its storage requirements without a significant loss of crucial details.

For consistency and convenience, all vectors in our approach are expressed in triple channels, namely the YUV color space, allowing for straightforward integration with existing color encoding and decoding schemes. This choice of representation facilitates seamless compatibility with various encoding and decoding frameworks, promoting ease of use and interoperability.

A sparse convolution is a technique that leverages the sparsity inherent in data to optimize convolutional operations. In a conventional convolution, computations are performed on every element, even if a significant portion of them are zero. This approach incurs unnecessary calculations and increases computational costs, particularly when dealing with large datasets like point clouds. However, with sparse convolutions, only non-zero elements are selectively processed during convolutional operations. By directing computations toward the sparse, non-zero elements, the computational overhead associated with zero elements is bypassed, resulting in improved computational efficiency.

PCAC-GAN uses an autoencoder with sparse convolution layers to effectively extract features, while also considering computational costs and storage constraints. A GAN is then used to generate high-quality point cloud data. As  Fig.~1 shows, the voxelized point cloud data is first expressed using a sparse tensor and then sent to the encoder. To accommodate sparse tensors, we use sparse convolutional layers and ReLU activation layers in the encoder. The quantizer then converts the encoder's output into a binary file for compression and transmission to the decoder. The decoder is essentially a generator framework that helps train the compression model by improving performance alternately. Specifically, the generator generates reconstructed point clouds that are close to the real point cloud based on the input data stream. The discriminator then decides whether the input point cloud stems from the generator's output or the original point cloud according to the designed loss function. During the training stage, the generator consistently attempts to persuade the discriminator, while the discriminator adjusts and optimizes the generator's neural network parameters based on the judgment results. This process continues until the generator produces texture and chroma as close as possible to the original point cloud. We next detail each model in our framework.

The encoder network comprises three sparse convolution layers and two ReLU activation layers. The convolution layers have kernel sizes of 9, 5, and 5, respectively, and except for the last three layers, have 128 channels to support high-dimensional feature embedding. The output of the encoder comprises the extracted features $\hat{\textbf{Y}}$, which are one-eighth of the geometric size of the original attribute $X$. After receiving the compressed point cloud data from the encoder, the generator’s objective is to reconstruct the original point cloud. To achieve this, the generator uses a network structure mirroring that of the encoder but replaces the sparse convolution layers with transposed sparse convolution layers. The generator produces the reconstructed point cloud by processing the feature map generated by the sparse convolution network through the sparse tensor with occupied voxel processing. By using this strategy, the generator is able to create a reconstructed point cloud that is nearly indistinguishable from the original.

The discriminator is a classification network that distinguishes the generated point cloud from the original.  It shares the same structure as the encoder for feature extraction. After activation, a dropout layer randomly sets a portion of the activations to zero, which helps prevent overfitting. The flattened layer then produces one-dimensional data, and the dense layer is used for classification.

The cost function for rate-distortion optimization is
\begin{equation}
  L = \lambda D + R,
  \label{eq5}
\end{equation}
where $D$ is the distortion, $R$ is the rate measured by the number of bits, and $\lambda$ is a Lagrange multiplier used to balance the relative importance of the distortion and the rate. To train the entire network, D is calculated as
\begin{equation}
  D=\phi_\text{adv} L_\text{adv} + \phi_\text{dec} L_\text{dec},
\end{equation}
where $L_\text{adv}$ is the adversarial loss, $L_\text{dec}$ is the decompression loss; $\phi_\text{adv}$ and $\phi_\text{dec}$ are parameters used to balance the adversarial and decompression losses.

As in the classical GAN framework, we define the adversarial loss $L_\text{adv}$  as 
\begin{equation}
L_{\text{adv}} = -\mathbb{E}\left[\log \mathcal{D}(\mathbf{m})\right] - \mathbb{E}\left[\log(1 - \mathcal{D}({\mathbf{G}}(\hat{\mathbf{n}})))\right],
\end{equation}
where $\mathcal{D}$ is the discriminator, G is the generator, \textbf{m} is sampled from the real data distribution  $p_r(m)$, \textbf{$\hat{n}$} is the compressed output data, and $\mathbb{E}[\cdot]$ is the expectation operator.

For the decompression loss, our method uses a classification-based decoding approach where the sparse tensor is classified as either occupied or unoccupied. Point clouds are typically highly sparse, with the majority of voxels being empty. Consequently, the sparse tensor that represents the voxelized point cloud contains very few non-zero values. This sparsity poses a challenge for conventional networks, as they are unable to effectively learn the 3D point cloud model. To achieve a better balance between occupied and unoccupied voxels, we use the $\sigma\text{-balanced}$ focal loss \cite{42}
\begin{equation}
L_\text{dec} = \sum_{w} -\bm{\sigma_w} (1 - \bm{p_w^t})^{\bm{\xi}} \log \bm{p_w^t},
\end{equation}
where $w$ is a voxel, \(\mathbf{\sigma_w}\) is a weight associated with $w$, \(\mathbf{p_{w}^{t}}\) is the probability that voxel $w$ belongs to category $t$, and \(\mathbf{\xi}\) is a focusing parameter used to regulate the rate of decrease in weight for easy samples. Here the sum is taken over all voxels; \(\mathbf{\sigma_w}\) is adjusted according to the color attributes of the voxel. The logarithm of \(\mathbf{p_{w}^{t}}\) is used because it is more sensitive to differences in low probabilities and less sensitive to differences in high probabilities than \(\mathbf{p_{w}^{t}}\). The use of focal loss helps the network to focus on areas in the point cloud that are more challenging to compress. This is achieved by reducing the relative loss in areas that are either already well compressed or are easy for the network to handle, such as homogeneous regions. This aids in improving the overall compression quality by not excessively penalizing simpler areas.

\begin{figure}[!t]
\centering
\includegraphics[width=\columnwidth]{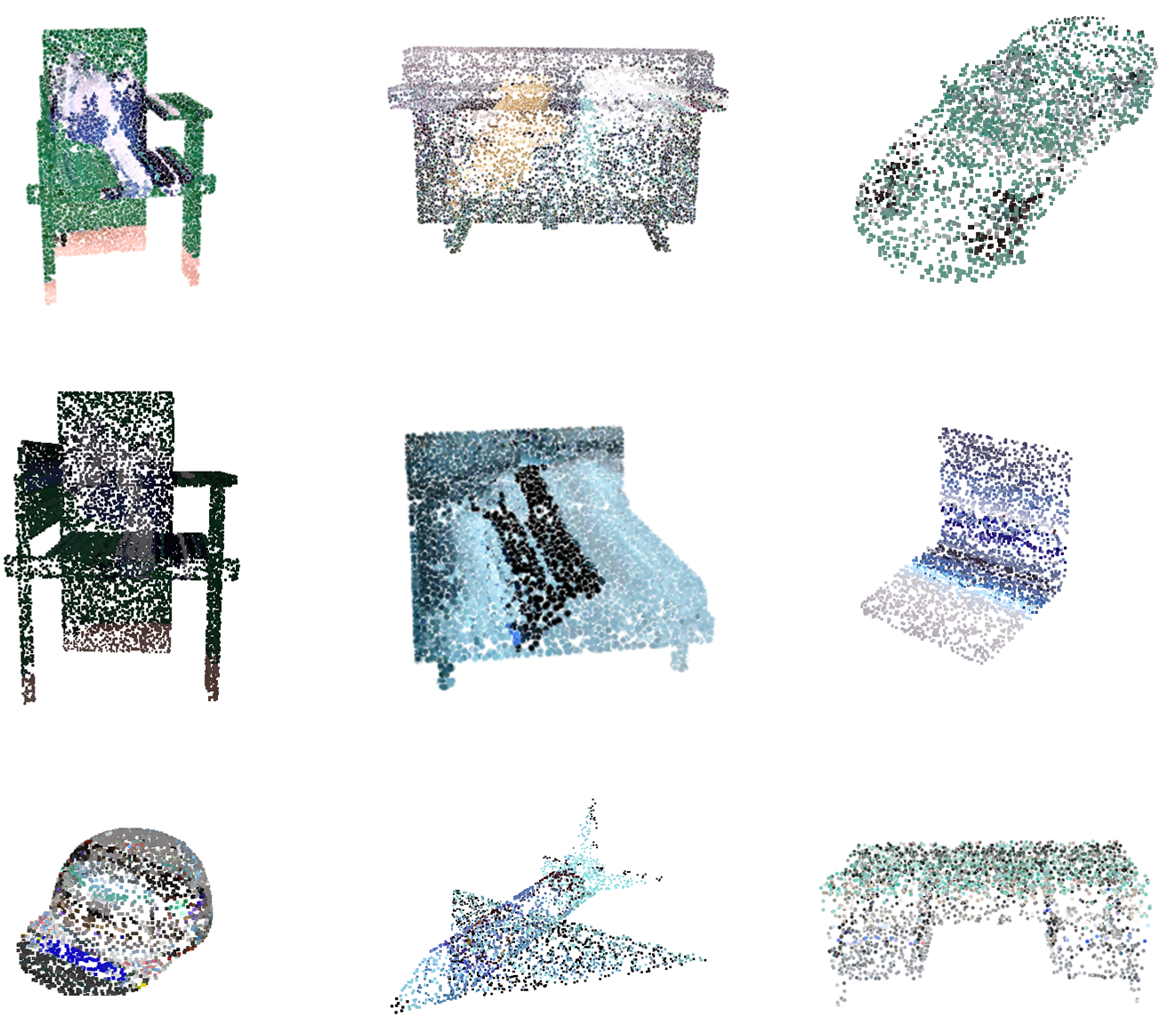}
\caption{Examples from the training dataset.}
\label{fig2}
\end{figure}

\section{Experimental Results}
Our implementation uses the PyTorch library, together with the Minkowski Engine \cite{49}, an Intel Xeon Gold6148 CPU with a base frequency of 2.40 GHz, 32 GB of RAM, and an NVIDIA GeForce RTX 4090 GPU.

\subsection{Model Training}
To build our datasets, we followed the approaches used in \cite{35,26}, which leverage two popular datasets,  ShapeNet and COCO.

First, we densely sampled points on the meshes provided by ShapeNet. These points were then subjected to random rotations and their coordinates were quantized to 8-bit integers. For the color attributes, we projected randomly selected images from the COCO dataset onto the corresponding points. Fig.~\ref{fig2} show some examples from the training dataset. Through this process, we generated 8,500 samples, which were subsequently divided into two distinct sets: 6,000 samples for training purposes and 2,500 samples for testing and evaluation. To mitigate the risk of overfitting, we incorporated an additional 11,000 samples from the ModelNet40 dataset into our training dataset, while maintaining the same training-testing ratio as before.

\begin{table*}[t]
  \centering
  \caption{BD-BR(\%) and BD-PSNR (dB) for our proposed method (codec under test) relative to SparsePCAC, TMC13v23, and TMC13v6 (reference codecs). The BD values indicate an increase (+) or decrease (-) in PSNR or BR of the  codec under test compared to the reference codec}
  \resizebox{\linewidth}{!}{
    \begin{tabular}{p{4.19em}|cc|cc|cc|cc|cc|cc}
    \specialrule{0.8pt}{0pt}{0.8pt}
    \multirow{3}[6]{*}{Point cloud} & \multicolumn{4}{c|}{SparsePCAC} & \multicolumn{4}{c|}{TMC13v23} & \multicolumn{4}{c}{TMC13v6} \bigstrut\\
\cline{2-13}    \multicolumn{1}{c|}{} & \multicolumn{2}{c|}{BD-BR (\%)} & \multicolumn{2}{c|}{BD-PSNR (dB)} & \multicolumn{2}{c|}{BD-BR (\%)} & \multicolumn{2}{c|}{BD-PSNR (dB)} & \multicolumn{2}{c|}{BD-BR (\%)} & \multicolumn{2}{c}{BD-PSNR (dB)} \bigstrut\\
\cline{2-13}    \multicolumn{1}{c|}{} & \multicolumn{1}{c}{Y} & \multicolumn{1}{c|}{YUV} & \multicolumn{1}{c}{Y} & \multicolumn{1}{c|}{YUV} & \multicolumn{1}{c}{Y} & \multicolumn{1}{c|}{YUV} & \multicolumn{1}{c}{Y} & \multicolumn{1}{c|}{YUV} & \multicolumn{1}{c}{Y} & \multicolumn{1}{c|}{YUV} & \multicolumn{1}{c}{Y} & \multicolumn{1}{c}{YUV} \bigstrut\\
    \hline
    Longdress & -10.01 & -10.35 & +0.41  & +0.27  & +27    & +9    & -0.54 & -0.67 & -22   & -25   & +1.77  & +1.51 \\
    Soldier & -2.56 & -2.13 & +0.16  & +0.09  & +38    & +47   & -0.58 & -1.19  & -15   & -16   & +1.18  & +1.16 \\
    Loot  & -1.73 & -1.87 & +0.29  & +0.16  & +32   & +45    & -0.52 & -0.34 & -17   & -19   & +1.01  & +0.82 \\
    Red\&black & -6.24 & -3.97 & +0.42  & +0.39  & +29    & +39   & -0.40 & -0.33  & -27   & -17   & +1.67  & +1.54 \\
    Queen & -4.65 & -4.15 & +0.31  & +0.23  & +37    & +44    & -0.39 & -0.50  & -20   & -13   & +1.69  & +1.58 \\
    Ford  & -3.65 & -3.33 & +0.23  & +0.14  & +44    & +36    & -0.31 & -0.51 & -15   & -11   & +1.37  & +1.21 \\
    Facade & -0.43 & -0.05 & +0.07  & +0.04  & +49    & +54    & -0.58 & -0.68 & -17   & -16   & +1.24  & +1.18 \\
    Dancer & -4.65 & -2.81 & +0.26  & +0.17  & +47    & +39    & -0.39 & -0.56  & -16   & -13   & +1.32  & +1.21 \\
    Boxer & -1.29 & -0.54 & +0.08  & +0.04  & +34    & +48    & -0.53 & -0.60  & -16   & -12   & +1.26  & +1.4 \\
    Andrew & -8.76 & -3.34 & +0.4   & +0.2   & +41    & +55    & -0.77 & -1.24 & -26   & -21   & +1.73  & +1.46 \\
    \hline
    \textbf{Average} & \textbf{-4.4} & \textbf{-3.25} & \textbf{+0.26} & \textbf{+0.17} & +38    & +43    & -0.50 & -0.66 & \textbf{-19} & \textbf{-16} & \textbf{+1.42} & \textbf{+1.31} \bigstrut\\
   \specialrule{0.8pt}{0pt}{0.8pt}
   \end{tabular}}
  \label{tab1}
\end{table*}%

\begin{figure}[t]
\centering
\includegraphics[width=0.8\columnwidth]{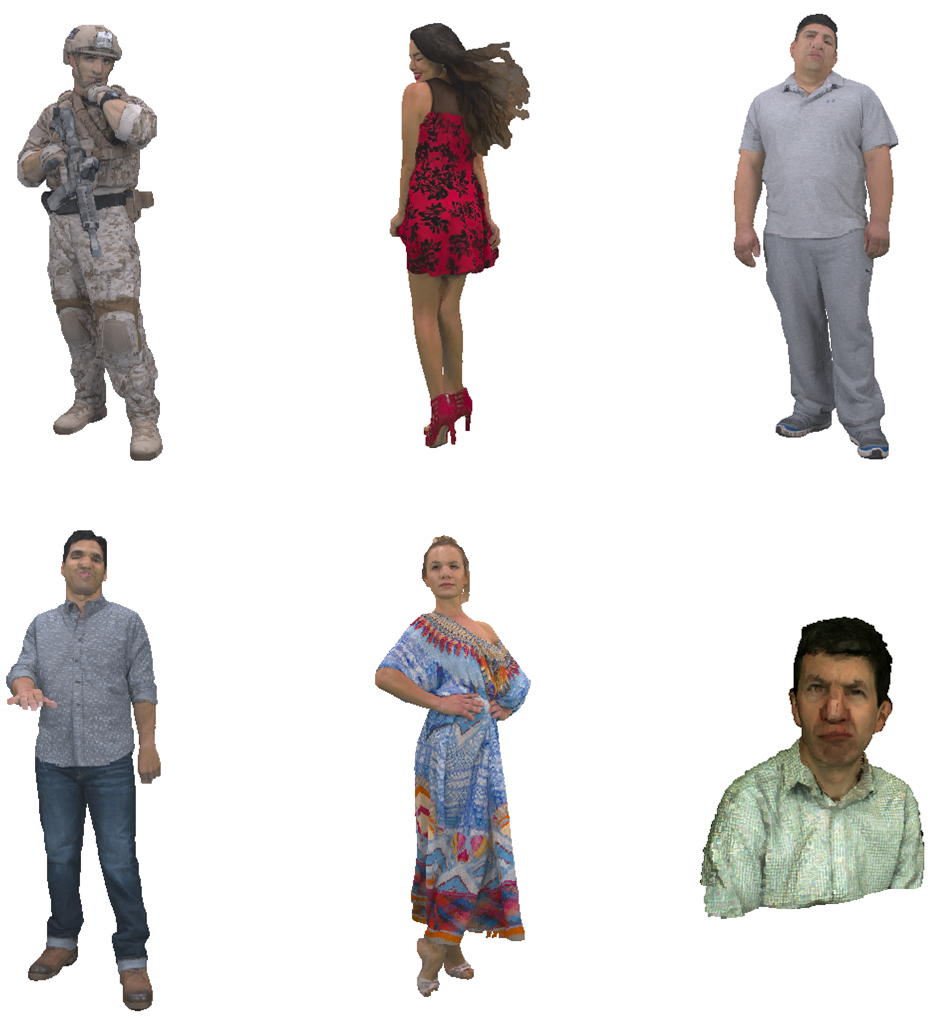}
\caption{Samples from the test dataset. Above: Soldier, RedandBlack, Boxer. Below: Loot, Longdress, Andrew.}
\label{fig3}
\end{figure}
\begin{figure*}[t]
  \centering
  \includegraphics[width=\textwidth]{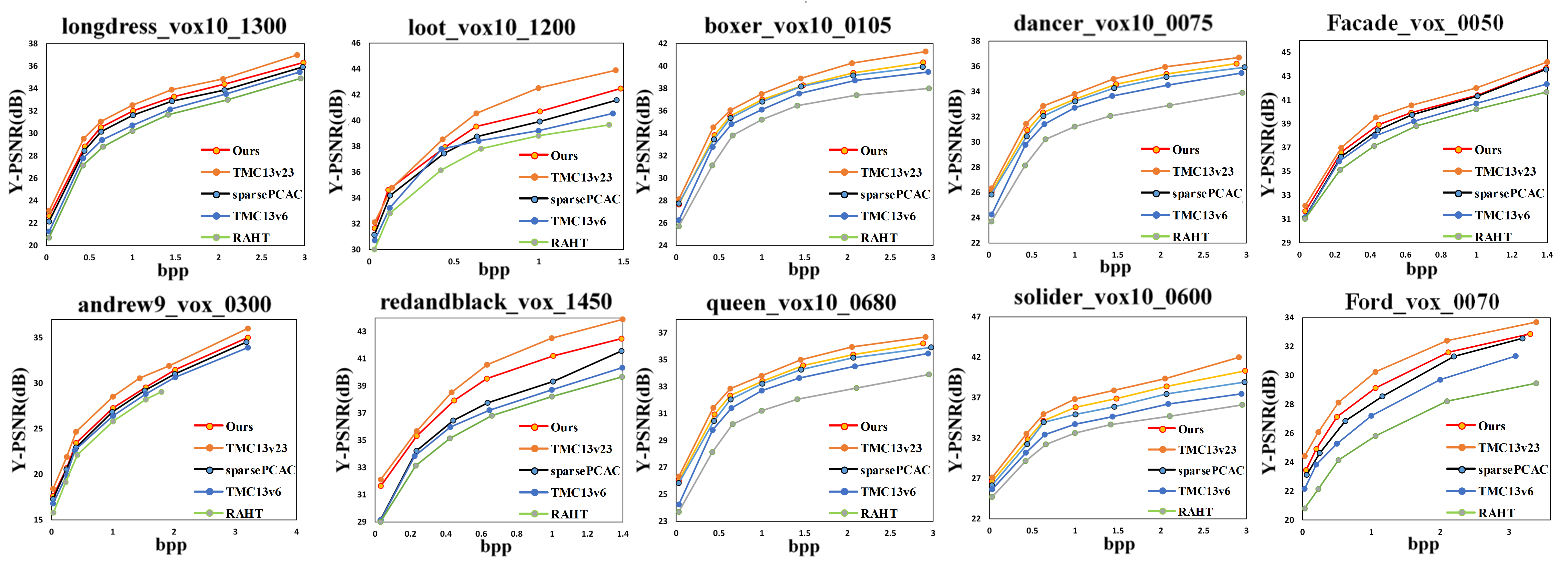}
  \caption{Y-PSNR versus bitrate in bits per input point (bpip).}
  \label{fig4}
\end{figure*}

\begin{figure*}[t]
  \centering
  \includegraphics[width=\textwidth]{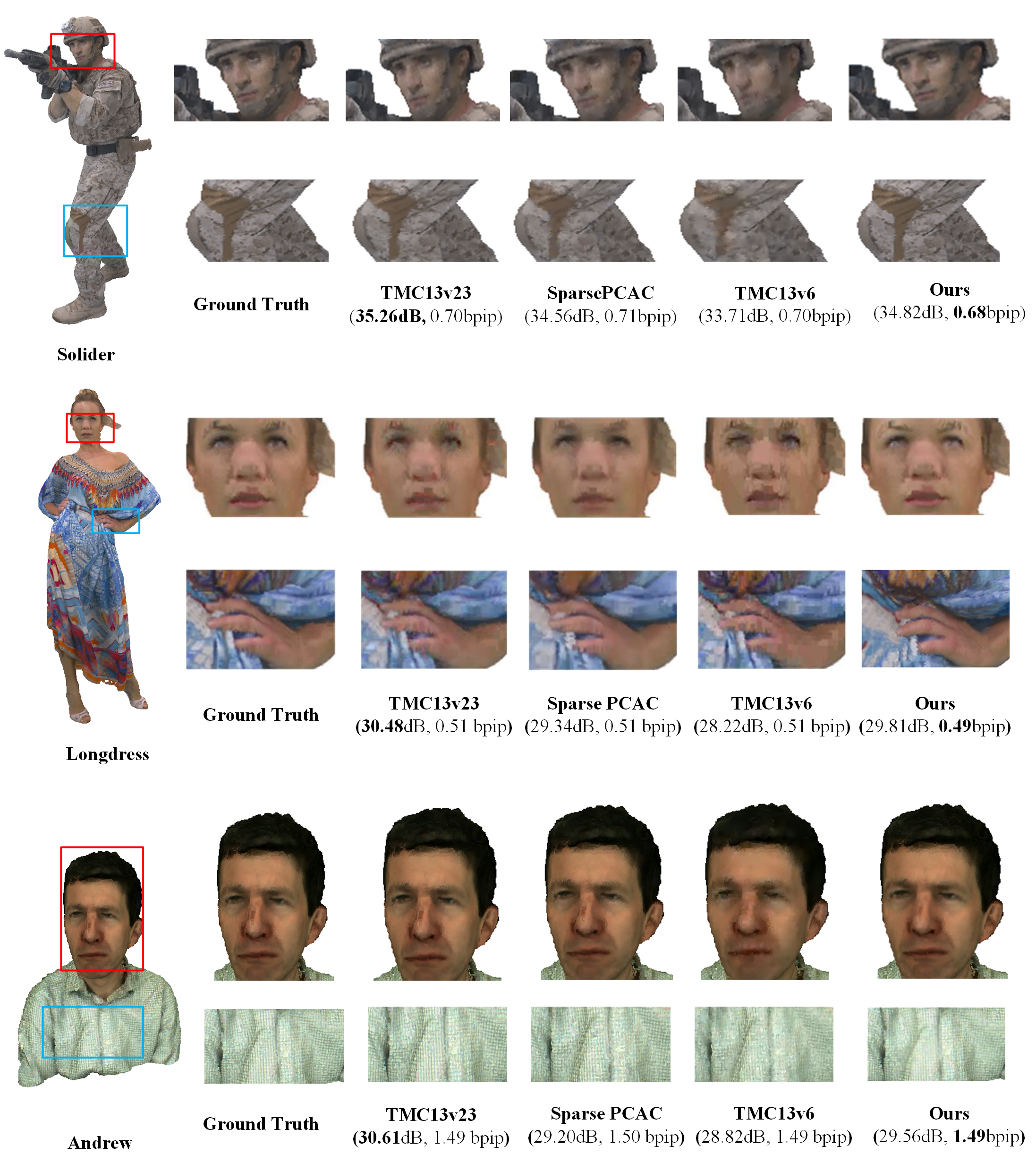}
  \caption{Reconstructed examples for the proposed method (Ours), SparsePCAC, TMC13v23, and TMC13v6.}
  \label{fig5}
\end{figure*}
The training loss function is given in Eq.~\eqref{eq5}. To generate bitstreams with various bit rates, we varied the value of the parameter $\lambda$ in Eq.~\eqref{eq5}. Specifically, we trained seven models by setting $\lambda$ to seven values. We first trained a model with a high bit rate by setting $\lambda$ to 0.0125. Subsequently, we used this model to initialize the other models at lower bit rates. The learning rate was set to $10^{-5}$. The batch size was set to 100, and the model was trained for $2\times10^5$ iterations. Moreover, the parameters $\phi_\text{adv}$ and $\phi_\text{dec}$ were set to 0.4 and 0.6, respectively.

\subsection{Performance Evaluation}
To comprehensively evaluate the effectiveness of our method, we conducted experiments using well-established datasets and compared our results to state-of-the-art techniques. In our evaluation, we used the 8i Voxelized Full Bodies (8iVFB) dataset \cite{43}, which is commonly used in MPEG standardization activities. This dataset provides a diverse range of point cloud data, offering various body shapes and poses, making it ideal for comprehensive evaluation.

Additionally, we incorporated the Andrew point cloud into our evaluation. This point cloud belongs to the sparsely populated MVUB class \cite{44}. This particular dataset presents a unique challenge due to its sparse nature, allowing us to assess the performance of our method under different data characteristics and densities. A sample of the test dataset is shown in Fig.~\ref{fig3}.

We compared our method to three established techniques: TMC13v23, the latest G-PCC test model, TMC13v6 with improved entropy coding, and SparsePCAC \cite{26}, a state-of-the-art learning-based point cloud attribute compression method.

To evaluate   performance objectively, we followed standard practice by calculating the peak signal-to-noise ratio (PSNR) as a function of the bitrate for the Y, U, V, and YUV channels. We did so using the MPEG PCC pc-error tools \cite{45}. Additionally, we used the Bjøntegaard delta (BD)-PSNR and BD-bitrate (BR) to measure the average rate-distortion performance.

The BD results are presented in Table~\ref{tab1}, while Fig.~\ref{fig4} shows the rate-distortion curves. We also conducted an additional comparison to the results of the original RAHT implementation \cite{7} In Fig.~\ref{fig4}. Our method outperformed TMC13v6, with a 19\% reduction in BD-BR and a 1.42 dB increase in BD-PSNR in the Y channel. Furthermore, our method outperformed SparsePCAC by reducing  BD-BR by 4.40\% and increasing  BD-PSNR by 0.26 dB in the Y channel. While our method was still behind TMC13v23 in terms of objective performance, it outperformed it in terms of subjective visual quality, especially in the high-frequency components. Fig.~\ref{fig5} shows the reconstructed point clouds.  PCAC-GAN outperformed all other methods in recovering high-frequency features. For Longdress, for example, PCAC-GAN achieved a better reconstruction of the finger contours and texture details than TMC13v23. Compared to the other methods, PCAC-GAN produced smoother facial areas and avoided color noise in the lips area.

\subsection{Ablation Study}

In this section, we describe four ablation studies. The aim of the studies was to analyze the impact of the discriminator, the sparse convolutions, and AVRPM. In the first study, we removed the discriminator from the PCAC-GAN architecture and retrained the model without modifying anything else. The results in Table~\ref{tab2} confirm the crucial role of the adversarial network in our method.

\begin{table}[t!]

  \caption{BD-PSNR (dB) for the YUV compound component. The reference codec is PCAC-GAN without discriminator and the test codec is PCAC-GAN with discriminator}
  
   \centering
   
  \label{tab2}
  \setlength{\tabcolsep}{0.7mm}
  \begin{tabular}{cccccc}
    \toprule
    Point cloud& Longdress& Soldier& Redandblack& Andrew& Average\\
    \midrule
    BD-PNSR (dB) & +2.31& +1.98& +2.24& +2.54& \textbf{+2.27}\\
  \bottomrule

  \centering
  
\end{tabular}
\end{table}

To demonstrate the benefits of sparse convolutions in PCAC-GAN, we created two additional experimental groups with identical settings for training data, network architecture, and hyperparameters. One group used sparse convolutions as proposed in this paper, while the other group used a 3D convolution instead. The experimental results presented in Table~\ref{tab3} demonstrate that the use of sparse convolutions reduced the bitrate and the time cost.
\begin{table}[t!]
  \caption{BD-BR (\%) of the YUV compound component, encoding time (\%) and decoding time (\%) comparisons between PCAC-GAN with sparse convolution and with standard 3D convolution. In the table,  “-” indicates a decrease}
  \label{tab3}
  \centering
  \setlength{\tabcolsep}{0.8mm}{
  \begin{tabular}{lccccc}
    \toprule
    Point cloud& Longdress& Soldier& Redandblack& Andrew& Average\\
    \midrule
    BD-BR (\%) & -4.23& -1.52& -3.51& -6.25& \textbf{-3.88}\\
    EncTime (\%) &-23 &-24 &-21 &-13 &\textbf{-20.25}\\
    DecTime (\%) &-18 & -12& -11& -11 &\textbf{-13}\\
  \bottomrule
\end{tabular}}
\end{table}

Lastly, to validate the benefits gained from AVRPM, we uniformly voxelized the raw point cloud into voxels with a resolution of $16\times 16 \times 16$ and proceeded with  encoding and decoding. The results in Table~\ref{tab4} show that voxelization using AVRPM effectively preserves point cloud details, thus improving the performance. Additionally, to assess the impact of AVPRM on encoding time, we also examined the encoding time implications with and without the use of this module. The results  in Table~\ref{tab5} show the influence of this module on encoding time, revealing that the increase in encoding time remains within acceptable limits.

\begin{table}[t!]
  \caption{BD-BR (\%) and BD-PSNR (dB) for the YUV compound component. The reference codec is PCAC-GAN without AVRPM and the test codec is PCAC-GAN with AVRPM}

  \centering
  
  \setlength{\tabcolsep}{0.7mm}{
  \begin{tabular}{lccccc}
    \toprule
    Point cloud& Longdress& Soldier& Redandblack& Andrew& Average\\
    \midrule
    BD-BR (\%) & -0.45& -0.23& -0.51& -0.53& \textbf{-0.43}\\
    BD-PSNR (dB) &-+0.1 &+0.06 &+0.13 &-+0.12 &\textbf{+0.10}\\
  \bottomrule

  \centering
  
\end{tabular}}
\label{tab4}
\end{table}

\begin{table}[t!]
\renewcommand{\arraystretch}{1}
\caption{Encoding time comparison between PCAC-GAN with AVPRM and PCAC-GAN without AVPRM}
\centering
\setlength{\tabcolsep}{0.8mm}{
\begin{tabular}{l|r|rl} 
\cline{1-3}
\multirow{2}{*}{Sequence } & PCAC-GAN           & PCAC-GAN w/o AVPRM &   \\ 
\cline{2-3}
                                             & Encoding Time (s) & Encoding Time (s) &   \\ 
\cline{1-3}
Longdress                  & 90.78              & 80.46              &   \\
Soldier                    & 79.71              & 62.12              &   \\
Loot                       & 83.24              & 74.87              &   \\
Red\&black                 & 83.33              & 73.26              &   \\
Queen                      & 100.07             & 84.29              &   \\
Ford                       & 87.69              & 62.76              &   \\
Façade                     & 217.07             & 175.61             &   \\
Dancer                     & 129.6              & 102.6              &   \\
Boxer                      & 87.29              & 65.59              &   \\
Andrew                     & 83.19              & 68.26              &   \\ 
\cline{1-3}
Average                    & 104.19             & 85.08              &   \\
\cline{1-3}
\end{tabular}}
\label{tab5}
\end{table}

\section{Conclusion}
We proposed the first GAN framework for the compression of point cloud attributes. The decoder uses multiscale transposed sparse convolution connections to ensure high-quality reconstruction of point cloud data at all bit rates. An adaptive voxel resolution partitioning module is designed to partition the point cloud into voxels with varying densities to preserve the data details. Our method outperformed SparsePCAC and TMC13v6 in terms of BDBR, BDPSNR, and subjective visual quality. While our method had inferior BD-BR and BD-PSNR performance compared to TMC13v23, it offered a better visual reconstruction quality. However, it should be noted that making a direct comparison with TMC13v23 is not entirely fair since our method uses a generative approach, placing it at a disadvantage in the comparison. This disadvantage arises due to the inherent differences in complexity and objectives between generative methods and conventional coding methods, which may affect both compression efficiency and assessment of generated content quality. We anticipate that with further improvements to filtering and cross-scale correlation for prediction, our method will outperform TMC13v23.


\begin{thebibliography}{99}

\bibitem{1} de Oliveira Rente, Paulo and Brites, Catarina and Ascenso, \emph{et al.}, ``Graph-based static 3D point clouds geometry coding," in \emph{IEEE Transactions on Multimedia}, \textbf{21}(2), 284-299(2018).

\bibitem{2} Garcia, Diogo C and Fonseca, Tiago A and Ferreira, \emph{et al.}, ``Geometry coding for dynamic voxelized point clouds using octrees and multiple contexts," in \emph{IEEE Transactions on Image Processing}, \textbf{29}, 313-322(2019).

\bibitem{3} Krivokuca, Maja and Chou, Philip A, \emph{et al.}, ``A volumetric approach to point cloud compression--part ii: Geometry compression," in \emph{IEEE Transactions on Image Processing}, \textbf{29}, 2217--2229(2019).

\bibitem{4} Chou, Philip A and Koroteev, Maxim and Krivokuca, Maja, \emph{et al.},``A volumetric approach to point cloud compression—Part I: Attribute compression," in \emph{IEEE Transactions on Image Processing}, \textbf{29}, 2203--2216(2010).
\bibitem{5} Gu, Shuai and Hou, Junhui and Zeng, \emph{et al.},``3D point cloud attribute compression using geometry-guided sparse representation," in \emph{IEEE Transactions on Image Processing}, \textbf{29},796--808(2019).
\bibitem{6} Xu, Yiqun and Hu, Wei and Wang, \emph{et al.},``Predictive generalized graph Fourier transform for attribute compression of dynamic point clouds," in \emph{IEEE Transactions on Circuits and Systems for Video Technology}, \textbf{31}, 1968--1982(2020).
\bibitem{7} De Queiroz, Ricardo L and Chou, Philip A, \emph{et al.},``Compression of 3D point clouds using a region-adaptive hierarchical transform," in \emph{IEEE Transactions on Image Processing}, \textbf{25}, 3947--3956(2016).
\bibitem{8} Xu, Yiqun and Hu, Wei and Wang, \emph{et al.},``Cluster-based point cloud coding with normal weighted graph fourier transform," in \emph{IEEE International Conference on Acoustics, Speech and Signal Processing (ICASSP)}, \textbf{29}, 1753--1757(2018).

\bibitem{9} Zhang, Cha and Florencio, \emph{et al.},``Point cloud attribute compression with graph transform," in \emph{IEEE International Conference on Image Processing (ICIP)}, \textbf{29},2066--2070(2014).

\bibitem{10} Kathariya, Birendra and Zakharchenko, \emph{et al.},``Level-of-detail generation using binary-tree for lifting scheme in LiDAR point cloud attributes coding," in \emph{2019 data compression conference (DCC)}, \textbf{29}, 580--580(2019).

\bibitem{11} Mammou, Khaled and Tourapis, \emph{et al.},``Lifting scheme for lossy attribute encoding in TMC1," in \emph{Document ISO/IEC JTC1/SC29/WG11 m42640, San Diego, CA, US},(2018).

\bibitem{12} Mammou, Khaled and Tourapis, \emph{et al.},``Video-based and hierarchical approaches point cloud compression," in \emph{Document ISO/IEC JTC1/SC29/WG11 m41649, Macau, China}, (2017).

\bibitem{13} Li, Li and Li, Zhu and Liu, Shan, \emph{et al.},``Efficient projected frame padding for video-based point cloud compression," in \emph{IEEE Transactions on Multimedia}, \textbf{23}, 2806--2819(2020).

\bibitem{14} Mekuria, Rufael and Blom, \emph{et al.},``Design, implementation, and evaluation of a point cloud codec for tele-immersive video," in \emph{IEEE Transactions on Circuits and Systems for Video Technology}, \textbf{27}, 828--842(2016).

\bibitem{15} Wallace, Gregory K, \emph{et al.},``The JPEG still picture compression standard," in \emph{Communications of the ACM}, \textbf{34}, 30--44(1991).

\bibitem{16} Sullivan, Gary J and Ohm, \emph{et al.},``Overview of the high efficiency video coding (HEVC) standard," in \emph{IEEE Transactions on circuits and systems for video technology}, \textbf{22}, 1649--1668(2012).

\bibitem{17} Balle, Johannes and Minnen, David, \emph{et al.},``Variational image compression with a scale hyperprior," in \emph{arXiv preprint arXiv:1802.01436}, (2018).

\bibitem{18} Chen, Tong and Liu, \emph{et al.},``End-to-end learnt image compression via non-local attention optimization and improved context modeling," in \emph{IEEE Transactions on Image Processing}, \textbf{30}, 3179--3191(2021).

\bibitem{19} Minnen, David and Balle, \emph{et al.},``Joint autoregressive and hierarchical priors for learned image compression," in \emph{Advances in neural information processing systems}, \textbf{31}, (2018).

\bibitem{20} Guarda, Andre FR and Rodrigues, \emph{et al.},``Adaptive deep learning-based point cloud geometry coding," in \emph{IEEE Journal of Selected Topics in Signal Processing}, \textbf{15}, 415--430(2020).

\bibitem{21} Nguyen, Dat Thanh and Quach, \emph{et al.},``Lossless coding of point cloud geometry using a deep generative model," in \emph{IEEE Transactions on Circuits and Systems for Video Technology}, \textbf{31}, 4617--4629(2021).

\bibitem{22} Quach, Maurice and Valenzise, \emph{et al.},``Improved deep point cloud geometry compression," in \emph{IEEE 22nd International Workshop on Multimedia Signal Processing (MMSP)}, \textbf{29}, 1--6(2020).

\bibitem{23} Huang, Lila and Wang, \emph{et al.},``Octsqueeze: Octree-structured entropy model for lidar compression," in \emph{Proceedings of the IEEE/CVF conference on computer vision and pattern recognition}, 1313--1323(2020).

\bibitem{24} Que, Zizheng and Lu, Guo, \emph{et al.},``Voxelcontext-net: An octree based framework for point cloud compression," in \emph{Proceedings of the IEEE/CVF Conference on Computer Vision and Pattern Recognition}, 6042--6051(2021).

\bibitem{25} Wang, Jianqiang and Ding, \emph{et al.},``Multiscale point cloud geometry compression," in \emph{2021 Data Compression Conference (DCC)}, 73--82(2021).

\bibitem{26} Wang, Jianqiang and Ma, Zhan, \emph{et al.},``Sparse tensor-based point cloud attribute compression," in \emph{EEE 5th International Conference on Multimedia Information Processing and Retrieval (MIPR)}, 59--64(2022).

\bibitem{27} ``G-PCC codec description v23," in \emph{ISO/IEC JTC 1/SC 29/WG 7}.

\bibitem{28} Sheng, Xihua and Li, \emph{et al.},``Deep-pcac: An end-to-end deep lossy compression framework for point cloud attributes," in \emph{IEEE Transactions on Multimedia}, \textbf{24}, 2617--2632(2021).

\bibitem{29} Quach, Maurice and Valenzise,, \emph{et al.},``Folding-based compression of point cloud attributes," in \emph{IEEE International Conference on Image Processing (ICIP)}, 3309--3313(2020).

\bibitem{30} Zhang, Sujun and Zhang, \emph{et al.},``A 3D Haar wavelet transform for point cloud attribute compression based on local surface analysis," in \emph{2019 Picture Coding Symposium (PCS)}, 1--5(2019).

\bibitem{31} Chen, Yueru and Wang, \emph{et al.},``A efficient predictive wavelet transform for LiDAR point cloud attribute compression," in \emph{IEEE International Conference on Visual Communications and Image Processing (VCIP)}, 1--5(2022).

\bibitem{32} Liu, Hao and Yuan, Hui, \emph{et al.},``A hybrid compression framework for color attributes of static 3D point clouds," in \emph{IEEE Transactions on Circuits and Systems for Video Technology}, \textbf{32}, 1564--1577(2021).

\bibitem{33} Cohen, Robert A and Tian, \emph{et al.},``Attribute compression for sparse point clouds using graph transforms," in \emph{2016 IEEE International Conference on Image Processing (ICIP)}, 1374--1378(2016).

\bibitem{34} Song, Fei and Li, Ge, \emph{et al.},``Fine-grained correlation representation for graph-based point cloud attribute compression," in \emph{2022 IEEE International Conference on Multimedia and Expo (ICME)}, 1--6(2022).

\bibitem{35} Cohen, Robert A and Tian, \emph{et al.},``Point cloud attribute compression using 3-D intra prediction and shape-adaptive transforms," in \emph{2016 Data Compression Conference (DCC)}, 141--150(2016).

\bibitem{36} Chen, Yueru and Shao, \emph{et al.},``Point cloud attribute compression via successive subspace graph transform," in \emph{2020 IEEE International Conference on Visual Communications and Image Processing (VCIP)}, 66--69(2020).

\bibitem{37} Shao, Yiting and Zhang, \emph{et al.},``Attribute compression of 3D point clouds using Laplacian sparsity optimized graph transform," in \emph{2017 IEEE Visual Communications and Image Processing (VCIP)}, 1--4(2017).

\bibitem{38} De Queiroz, Ricardo L and Chou, Philip A, \emph{et al.},``Transform coding for point clouds using a Gaussian process model," in \emph{IEEE Transactions on Image Processing}, \textbf{26}, 3507--3517(2017).

\bibitem{39} Isik, Berivan and Chou, Philip A , \emph{et al.},``Lvac: Learned volumetric attribute compression for point clouds using coordinate based networks," in \emph{Frontiers in Signal Processing}, \textbf{2}, 1008812(2022).

\bibitem{40} Fang, Guangchi and Hu, \emph{et al.},``3dac: Learning attribute compression for point clouds," in \emph{Proceedings of the IEEE/CVF Conference on Computer Vision and Pattern Recognition}, 14819--14828(2022).

\bibitem{41} Ding, Dandan and Zhang, \emph{et al.},``CARNet: Compression Artifact Reduction for Point Cloud Attribute," in \emph{arXiv preprint arXiv:2209.08276},(2022).

\bibitem{42} Lin, Tsung-Yi and Goyal, \emph{et al.},``Focal loss for dense object detection," in \emph{Proceedings of the IEEE international conference on computer vision}, 2980--2988(2017).

\bibitem{43} d’Eon, Eugene and Harrison, \emph{et al.},``8i voxelized full bodies-a voxelized point cloud dataset," in \emph{ISO/IEC JTC1/SC29 Joint WG11/WG1 (MPEG/JPEG) input document WG11M40059/WG1M74006}, \textbf{7}, 11(2017).

\bibitem{44} Loop, Charles and Cai, \emph{et al.},``Microsoft voxelized upper bodies-a voxelized point cloud dataset," in \emph{ISO/IEC JTC1/SC29 Joint WG11/WG1 (MPEG/JPEG) input document m38673 M}, \textbf{72012}, 2016(2016).

\bibitem{45} Tian, Dong and Ochimizu, \emph{et al.},``Geometric distortion metrics for point cloud compression," in \emph{2017 IEEE International Conference on Image Processing (ICIP)}, 3460--3464(2017).

\bibitem{46} Liu, Zhijian and Tang, \emph{et al.},``Point-Voxel CNN for Efficient 3D Deep Learning," in \emph{Advances in Neural Information Processing Systems}, \textbf{32},(2019).

\bibitem{47} You, Kang and Gao, \emph{et al.},``Patch-Based Deep Autoencoder for Point Cloud Geometry Compression," in \emph{Association for Computing Machinery}, \textbf{30}, 7(2022).

\bibitem{48} Gao, Pan and Zhang, \emph{et al.},``Point Cloud Compression Based on Joint Optimization of Graph Transform and Entropy Coding for Efficient Data Broadcasting," in \emph{IEEE Transactions on Broadcasting}, \textbf{69}, 727-739(2023).

\bibitem{49} Choy, Christopher and Gwak, \emph{et al.},``4D Spatio-Temporal ConvNets: Minkowski Convolutional Neural Networks," in \emph{Proceedings of the IEEE/CVF Conference on Computer Vision and Pattern Recognition (CVPR)}, (2019).


\end{thebibliography}
\end{document}